\documentclass[12pt]{iopart}
\usepackage{iopams}  
\usepackage[utf8]{inputenc}
\usepackage{graphicx}
\usepackage{float,placeins}\usepackage[caption=false]{subfig}

\begin{document}

\title{A closure for the Master Equation starting from the Dynamic Cavity Method}
\author{Erik Aurell}
\address{KTH -- Royal Institute of Technology, AlbaNova University Center, SE-106 91 Stockholm, Sweden}%
\ead{eaurell@kth.se}

\author{David Machado Perez}
\address{
Group of Complex Systems and Statistical Physics. Department of Theoretical Physics,
Physics Faculty, University of Havana, Cuba}
\ead{davidmachadop@gmail.com}

\author{Roberto Mulet}
\address{
Group of Complex Systems and Statistical Physics. Department of Theoretical Physics,
Physics Faculty, University of Havana, Cuba}
\ead{roberto.mulet@gmail.com}

\begin{abstract}
We consider classical spin systems evolving in continuous time with interactions given by a locally tree-like graph. Several approximate analysis methods have earlier been reported based on the idea of Belief Propagation / cavity method. We introduce a new such method which can be derived in a more systematic manner, and which performs better on several important classes of problems.   
\end{abstract}

%
%
%
%
%

\section{Introduction}
\label{sec:introduction}

Problems across many scientific disciplines 
require understanding the non-equilibrium dynamics of many interacting variables. The variables could be spins in condensed-matter physics\cite{Onuki}, cells or reactions in biology \cite{Beard}, neurons in neurosciences and machine learning \cite{Hertz}, etc. In all these cases the mathematical formulation is almost always the same. Given $N$ variables $\underline{\sigma}=\{\sigma_1, \cdots, \sigma_N\}$ one needs, in principle, to solve the master equation:

  \begin{equation}
    \frac{d P(\underline{\sigma})}{dt} = - \sum_{i=1}^N  [ r_i( \underline{\sigma} \rightarrow \underline{\sigma'}) P(\underline{\sigma}) - r_i( \underline{\sigma'} \rightarrow \underline{\sigma}) P(\underline{\sigma'})]
    \label{eq:Master}
  \end{equation}
  where $P(\underline{\sigma})$ is the probability of the configuration $\underline{\sigma}$ and $r_i( \underline{\sigma'} \rightarrow \underline{\sigma})$ defines the transition rates from configuration $\underline{\sigma'}$ to $\underline{\sigma}$.

  Equation (\ref{eq:Master}) is a Markovian first order differential equation. Although it is compact and formally simple, it implies the tracking in time 
  the probabilities of $2^N$ discrete states, a daunting task that can be done only for very small systems. 
  A large amount of work has therefore been devoted to find approximate solutions or closure schemes able to provide an accurate yet computationally manageable description of this dynamics.

The transition rate $r_i$ of spin $i$ can in principle depend only on spin $i$ itself, on all the spins, or on spin $i$ and some set of neighbours $\partial i$. In the last case, which is the one considered here, the dependency sets of all the spins and how they are connected define a directed graph. In this graph there is a link $j\to i$ if and only if $j\in \partial i$; the dependency of $r_i$ of spin $i$ (if present) has to be taken into account separately.
It is straightforward to show that the probability of variable $\sigma_i$ only changes as
\begin{equation}
    \frac{d P(\sigma_i)}{dt} = - \sum_{i=1}^N  [ r_i(\sigma_i,\sigma_{\partial i}) P(\sigma_i,\sigma_{\partial i}) - r_i(-\sigma_i,\sigma_{\partial i}) P(-\sigma_i,\sigma_{\partial i})]
    \label{eq:OneMaster}
  \end{equation}
Although apparently simpler than (\ref{eq:Master}), equation (\ref{eq:OneMaster}) is not closed, to compute $P(\sigma_i)$ on the LHS one needs information about $P(\sigma_i, \sigma_{\partial i})$. The problem then reduces to find a proper closure scheme for equation  (\ref{eq:OneMaster}). The main goal of this work is to propose and test a new such scheme.

In parallel to (\ref{eq:Master}) and (\ref{eq:OneMaster}) one can also consider the analogous equations for discrete time. One example would be a time-discretization of (\ref{eq:Master}) and (\ref{eq:OneMaster}) with a finite time step $\Delta t$. The discrete time model allows also for other types of dynamics, but is nevertheless in a mathematical sense simpler. The history of a discrete-time variable over a finite time interval is a finite-dimensional variable, while the history of a continuous-time variable is infinite-dimensional. Many techniques introduced for discrete time further have only a trivial limit as $\Delta t$ tends to zero. In short, the continuous-time case is what is directly appropriate for most applications, and also needs additional treatment compared to the discrete case.  The work reported in this paper is another effort in this direction.

The paper is organized as follows. In section \ref{sec:dyncav} we review the Dynamic cavity method in the version introduced for discrete-time in \cite{Gino2015}. In Section \ref{sec:clousure} we introduce our new closure scheme for the continuous-time case. Then, Section \ref{sec:illustrations} shows the comparison of our solution with Monte Carlo simulations and with alternative closure schemes build on similar principles. Finally we present the conclusions of our work, they also summarize the main limitations of our approach and highlight possible paths for future developments.

\section{Dynamic cavity method}
\label{sec:dyncav}
\subsection{Definition, specificity and main problem}
\label{sec:definition}
The standard cavity method is a means to compute marginals of a Gibbs-Boltzmann distribution 
by exchanging messages~\cite{MezardMontanari}.
When it converges the cavity method is computationally
efficient requiring a number of operations polynomial in system size. The cavity method is exact when the interaction graph is a tree and asymptotically exact for many types of a locally tree-like interaction graphs.
As in this Letter we consider statistical inference we will not discuss the use
of dynamic cavity to retrodict the origin of epidemics and similar processes, for this
see \cite{Altarelli2014} and \cite{Lokhov2014}. To stay inside the sphere of physical problems
we will also not consider the recent use of dynamic cavity to model and predict the evolution 
of an epidemic \cite{Ortega2022,David2}.

We will consider a dynamics specified by an graph of the same locally tree-like type. Let the history of variable $i$ up to time
$t$ be $X^t_i$, and
let the value of variable
$i$ at time $t$ be $\sigma_i(t)$.
For an Ising variable we can formally define
$X^t_i=(\sigma_i(t_0),n,t_1,t_2,\ldots)$ 
where $\sigma_i(t_0)$ is the initial value of the spin,
$n$ is the number of jumps in the time interval 
$[t_0:t]$
and $t_1,t_2,\ldots$ are the set of spin flip times.

We will consider the setting where the one-time joint probability of all the variables 
satisfies a high-dimensional differential equation 
(Master equation) of the type
\begin{equation}
\label{eq:master-equation}
    \frac{d}{dt}P\left(\sigma_1,\ldots,\sigma_N,t\right) = \sum_{\sigma'_1,\ldots,\sigma'_N}
    \Gamma_{\mathbf{\sigma},\mathbf{\sigma}'} P\left(\sigma'_1,\ldots,\sigma'_N,t\right).
\end{equation}
The main idea is now to trade the high-dimensional probability $P\left(\sigma_1,\ldots,\sigma_N,t\right)$
with the infinite-dimensional probability $P(X^t_i)$. In so doing the goal is to arrive at an accurate and computationally effective description of the one-time marginal probabilities  $P_i\left(\sigma_i,t\right)$.
To proceed we first note that
the locally tree-like interaction graph describes the transition matrices
$\Gamma_{\mathbf{\sigma},\mathbf{\sigma}'}$ which
satisfy $\sum_{\mathbf{\sigma}}\Gamma_{\mathbf{\sigma},\mathbf{\sigma}'}=1$ for every
value of $\mathbf{\sigma}'$.
The joint probability over the histories of all the variables can then, up to technicalities,
be written 
\begin{equation}
\label{eq:transition-matrix-product}
    P^t(X^t_1,\ldots,X^t_N) = 
    \Gamma_{\mathbf{\sigma}(t),\mathbf{\sigma}(t-\epsilon)} \cdots \Gamma_{\mathbf{\sigma}(t_0+\epsilon),\mathbf{\sigma}(t_0)}
    \cdot P^0(\sigma_1,\ldots,\sigma_N,t_0).
\end{equation}
We either assume that the initial probability distribution $P^0$ is so far in the past that it does not matter, or that it only has the same dependencies as in
$\Gamma$. For instance, 
it can be factorized.

It follows from the Markovian nature of 
the Master equation
(\ref{eq:master-equation}) that the probabilities  
of different variables to flip in a short time interval $\Delta t=\epsilon$
are independent. This is in any case natural in the continuous-time limit 
where these probabilities are given by 
$\Delta t\cdot r_i$ where $r_i$ is the instantaneous flip rate of spin $i$.
For the Ising ferromagnet with Glauber dynamics,
which we show as a numerical example in
Fig.~\ref{fig.iCME}, the rates are
\begin{equation}
r_i(\sigma_i,\sigma_{\partial i})
= \alpha\frac{
e^{-\frac{J}{k_B T}\sum_{j\in\partial i}\sigma_i\sigma_j}}{
\sum_s e^{-\frac{J}{k_B T}\sum_{j\in\partial i}s \, \sigma_j}},
\end{equation}
where $\alpha$ is a constant of dimension inverse time and 
$J$ is the pairwise interaction energy.

The dependencies in the joint probability distribution (\ref{eq:transition-matrix-product})
includes effects of the type that if $k$ and $j$ are both
in the neighborhood of $i$ as given by the energy function, 
they are also related by the denominator in the expression for the
rate $r_i$. The total statistical dependencies in (\ref{eq:transition-matrix-product}) therefore include many local loops.
A systematic approach to resolve these loops is 
by graph expansion 
\cite{Altarelli2013,Gino2015}. This approach associates a pair of variables
$(X^t_i,X^t_j)$ to each link $(ij)$ in the original dependency
graph and imposes hard constraints $C_i$ that all variables of the type
$X^t_i$ in all links $(ij)$ take the same value.
The probability distribution (\ref{eq:transition-matrix-product})
can then equivalently be written 
\begin{equation}
\label{eq:transition-matrix-product-2}
    P^t\big(\{X^{t,(ij)}_i,X^{t,(ij)}_j\}\big) =
    \prod_i \Phi_i\big(X^{t}_i,\{X^{t,(ij)}_j\}_{j\in \partial i}\big)
    \prod_i C_i
\end{equation}
where the local loops have been resolved.
The first argument $X^{t}_i$ 
on the right hand of above can be any of the $X^{t,(ij)}_i$
as by the constraint $C_i$ they are all the same.
Using
the theory of Random Point Processes \cite{vanKampen}
the local weight functionals can further be written
\begin{eqnarray}
\Phi_i\left(X^{t}_i,X^{t}_{\partial i}\right)
&=& \prod_{s=1}^{n} r_i\left(\sigma_i(t_s),\sigma_{\partial i}(t_s)\right)
\cdot e^{-\int_{t_0}^{t_1} r_i\left(\sigma_i(\tau),\sigma_{\partial i}(\tau)\right)
\, d\tau} \nonumber \\
&& \qquad
\cdot \prod_{s=1}^{n} 
e^{-\int_{t_s}^{t_{s+1}} r_i\left(\sigma_i(\tau),\sigma_{\partial i}(\tau)\right) \, d\tau} \label{eq:phi_param}
\end{eqnarray}
where we recall that 
$X^{t}_i$ is defined by $n$, 
the number of jumps of spin $i$ in a time interval $[t_0,t_f]$, the initial 
spin state, and the jump times.
For given $n$ the last time ($t_{n+1}$) in above is $t_f$. 

After applying the graph expansion
the right-hand side of (\ref{eq:transition-matrix-product-2}) is like  
a Boltzmann weight in the standard cavity method 
with hard constraints, though over an infinite-dimensional space. 
The marginal probability over one history 
In the original formulation (\ref{eq:transition-matrix-product})
is defined as $P^t_i(X^t_i) = \sum_{\mathbf{X}_{\setminus i}} P^t(X^t_1,\ldots,X^t_N)$
and in the expanded graph we can first marginalize to the joint probability of the set $\{X^{t,(ij)}_i,X^{t,(ij)}_j\}_{j\in\partial i}$,
where all the $X^{t,(ij)}_i$ are the same due to the 
constraint $C_i$, and then marginalize separately over 
the $X^{t,(ij)}_j$.
The cavity (or Belief Propagation) output equation is
then
\begin{equation}
    P^t_i(X^t_i) = \sum_{X_{\partial i}^t} \Phi_i\left(X^t_i,X^t_{\partial i}\right)
    \, \prod_{j\in\partial i} \mu^t_{j\to (ji) }(X^t_j,X^t_i ).
       \label{eq:cavity-output-X}
\end{equation}
In above $\mu^t_{j\to (ji) }(X^t_j,X^t_i)$ (a message with two arguments)
follows from the graph expansion, 
and the egress node notation $(ji)$
indicates that these messages are 
actually passed around in the expanded graph.
The cavity (or BP) update equation is on the same level of abstraction
\begin{equation}
    \mu^t_{j\to (ji) }(X^t_j,X^t_i) =\sum_{X^t_{\partial j\setminus i}} \Phi_j\left(X^t_j,X^t_{\partial j}\right)  \prod_{k\in\partial j\setminus i} \mu_{k\to (kj)}(X^t_k,X^t_j).
    \label{eq:cavity-update-X}
\end{equation}
Equations (\ref{eq:cavity-output-X}) and (\ref{eq:cavity-update-X}) cannot be used as is since the   the argument is infinite-dimensional.
The equations need to be closed in a
suitable finite-dimensional subspace. Furthermore, for the cavity method to be 
computationally attractive, the subspace should have only one or at most a few spin degrees of freedom per node.

\section{Closure scheme}
\label{sec:clousure}
After deriving (\ref{eq:cavity-update-X}), the next step is to find a convenient parametrization for the histories. We will first briefly review the 
discrete-time setting, where a simple parametrization is to consider the values of the spins at different times. Each $X_i^t$ is then approximated by the values of the spins at different times, $X_i^t \approx (\sigma^t, \sigma^{t -\epsilon}, \cdots ,\sigma^0)$. To arrive at finite-dimensional messages one can then consider a closure on the last $n$ times, which means to take into account a memory of length $n\delta t$. This was the approach (for $n=2$) followed in \cite{Gino2015,Neri} when studying of the kinetic Ising model under synchronous update dynamics. A more advanced approach based on the matrix product expansion from quantum condensed matter theory was investigated in \cite{Barthel2018}.
Neither of these approaches extend to 
continuous time.

In a series of papers reviewed in \cite{Dominguez2020} a continuous-time closure 
was introduced leading to a cavity master equation.
Apart from the kinetic Ising model (pair-wise interactions)
this versatile approach has also been applied 
with good results to the ferromagnetic $p$-spin model under Glauber dynamics \cite{pspindyn}, and to the dynamics of a focused search algorithm to solve the random 3-SAT problem in a random graph \cite{ksatdyn}. 
The method has also generalized to provide master equations for the probability densities of any group of connected variables \cite{David2}. 
We will see that the systematic approach introduced
here will lead to additional terms, rendering the final formulae somewhat more symmetric and transparent.

In our new approach the starting point is the
final-time cavity marginalizations
\begin{equation}
    p_{i\to (ji)}\left(\sigma_i,\sigma_j\right)=\sum_{X^t_i:\sigma_i(t)=\sigma_i}
\sum_{X^t_j:\sigma_j(t)=\sigma_j}\mu_{i\to (ji)}\left(X^t_i,X^t_j\right)
\end{equation}
This is different from the earlier approach reviewed 
in \cite{Dominguez2020} where the starting point was
the single-site marginals
$p_{i}\left(\sigma_i\right)$.
The closure of the cavity update equations as master-equation-like
differential equations reads

\begin{eqnarray}
&\frac{d}{dt}{p}_{i\to (ij)}(\sigma_i,\sigma_j) =  \sum_{\sigma_{\partial i \setminus j}} \Big[ r_i(\sigma_i,\sigma_{\partial i})  \prod_{ k \in \partial i\setminus j} p_{k\to (ki)}(\sigma_k \mid \sigma_i) p_{i\to (ij)}(\sigma_i, \sigma_j) \nonumber\\ 
&\qquad- r_i(-\sigma_i,\sigma_{\partial i})  \prod_{ k \in \partial i\setminus j} p_{k\to (ki)}(\sigma_k \mid -\sigma_i) p_{i\to (ij)}(-\sigma_i, \sigma_j) \Big]\nonumber \\ 
&\qquad- r_j(\sigma_j,\sigma_i) p_{i\to (ij)}(\sigma_i,\sigma_j)
+ r_j(-\sigma_j,\sigma_j) p_{i\to (ij)}(\sigma_i,-\sigma_j).
\label{eq:iCME-update}
\end{eqnarray}
In the above the conditional probabilities in the cavity are defined as:
$p_{i\to (ij)}(\sigma_i \mid \sigma_j)=\frac{p_{i\to (ij)}(\sigma_i, \sigma_j)}{\sum_s p_{i\to (ij)}(s, \sigma_j)}$

Further,
$r_i(\sigma_i,\sigma_{\partial i})$ and $r_j(\sigma_j,\sigma_i)$ are the 
defined jump rates of spins
$i$ and $j$ in the cavity graph obtained by eliminating all neighbours of $j$ except $i$.
The rate $r_i$ hence depends on all neighbours of $i$ in the original graph, including $j$,
while the rate $r_j$ only depends on $i$ and $j$. 


The fundamental object of the cavity output equations
are analogously the final-time marginalizations $  P_i\left(\sigma_i\right)=\sum_{X^t_i:\sigma_i(t)=\sigma_i} P^t_i(X^t_i)$
and the differential equations substituting for (\ref{eq:cavity-output-X})
are
\begin{eqnarray}
\frac{d}{dt}{P}_i(\sigma_i) 
 &=  \sum_{ \sigma_{\partial i}}\Big[ r_i(\sigma_i,\sigma_{\partial i})  \prod_{ k \in \partial i\setminus j} p_{k\to (ki)}(\sigma_k | \sigma_i) P_i(\sigma_i) \nonumber \\ 
&\quad- r_i(-\sigma_i,\sigma_{\partial i})  \prod_{ k \in \partial i\setminus j} p_{k\to (ki)}(\sigma_k | -\sigma_i) P_i(-\sigma_i) \Big]. \label{eq:iCM-output}
\end{eqnarray}

In summary, we claim that the combination of equations (\ref{eq:iCME-update}) and (\ref{eq:iCM-output}) constitutes a proper closure scheme that should approximately describe the dynamics of the Master equation (\ref{eq:Master}) for continuous time and discrete variables.In the next section we will see how this expectation fares in numerical tests.
 
\section{Numerical illustrations}
\label{sec:illustrations}
\begin{figure}[h]
\centering
\subfloat[]{
\includegraphics[width=0.35\textwidth]{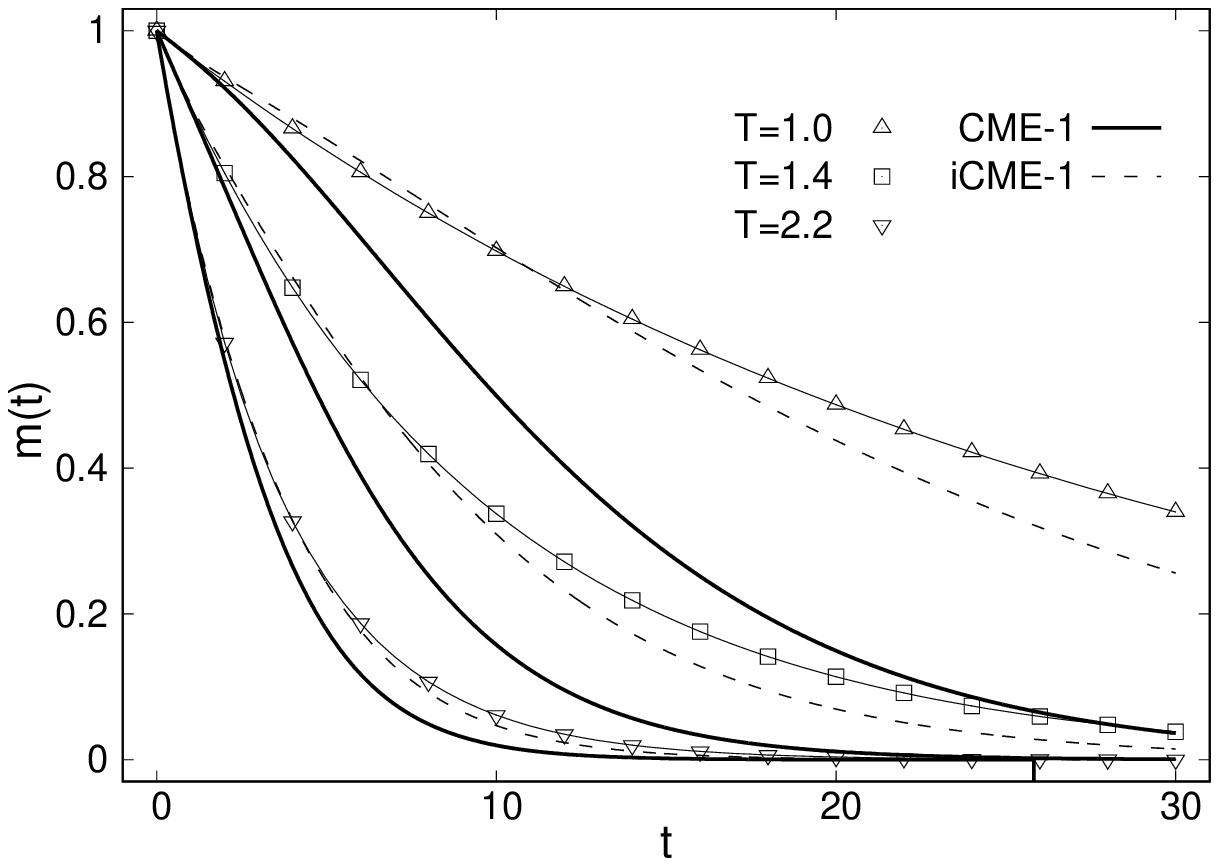}\label{fig:1D}
}
\subfloat[]{
\includegraphics[width=0.35\textwidth]{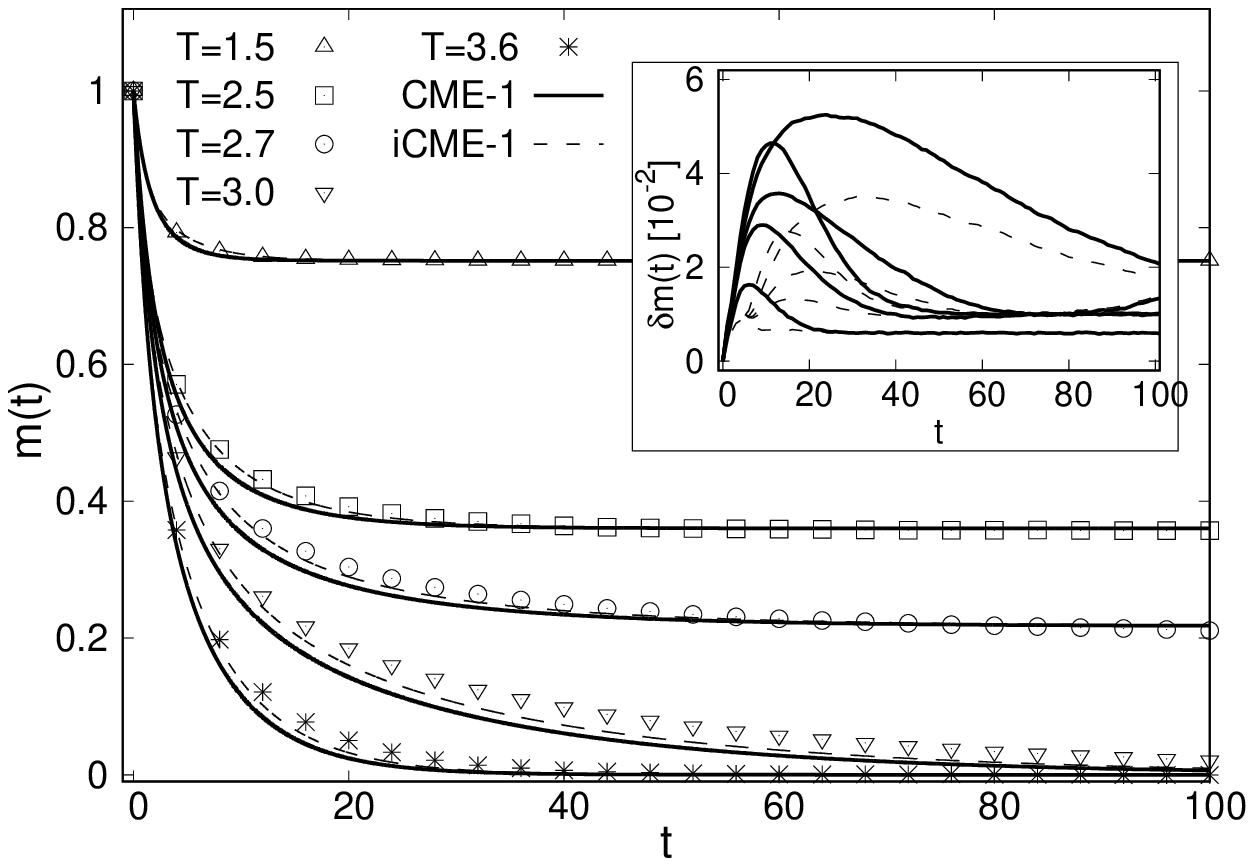}\label{fig:RGER}
}
\caption{Continuous-time dynamics of the Ising ferromagnet. Both panels compare the results of the earlier cavity closure \cite{CME-PRE} (thick continuous lines) and the new closure (dashed lines). 
In all calculations, an initially fully magnetized system evolves in time in contact with a heat bath at a given temperature. \textbf{a)} Magnetization of the one-dimensional Ising ferromagnet. Lines with points represent the exact Glauber's solution for the average magnetization \cite{Glauber63}. \textbf{b)} Magnetization in a single instance of an Erdos-Renyi graph with size $N=5000$ and average connectivity $c=3$.  
The main panel shows the time evolution of the system magnetization. Points are the averages of $s=10,000$ kinetic Monte Carlo simulations of the dynamics. The inserted graphic shows the mean square error $\delta m(t) = (N^{-1}  \sum_{i=1}^{N} (m_i^{DCAV}(t) - m_i^{MC}(t))^{2})^{1/2}$ on scale of order $10^{-2}$.}
\label{fig.iCME}
\end{figure}

In Fig.~\ref{fig.iCME} we show numerical results on the kinetic Ising model
obtained using (\ref{eq:iCME-update}) and (\ref{eq:iCM-output}); it can be checked 
that they improve on the earlier version of the continuous-time closure \cite{CME-PRE}. The left panel (Fig.~\ref{fig:1D}) contains results for the one-dimensional Ising ferromagnet, the exact solution of which was obtained by Glauber \cite{Glauber63} is represented with lines and points. In this simple model, the first version of the cavity closure is surprisingly far from the solution, while the new closure presented here gives considerably better results.

On the other hand, Fig.~\ref{fig:RGER} shows results on the Ising ferromagnet defined over an Erdos-Renyi graph. Both the original and the new closure give results that are very similar to the dynamics of the Monte Carlo simulations. Nevertheless, it is possible to appreciate a small improvement of the accuracy provided by the new approach. The local errors in the inserted graphic point to the same conclusion for all temperatures.

\begin{figure}[h]
\centering
\subfloat[]{
\includegraphics[width=0.35\textwidth]{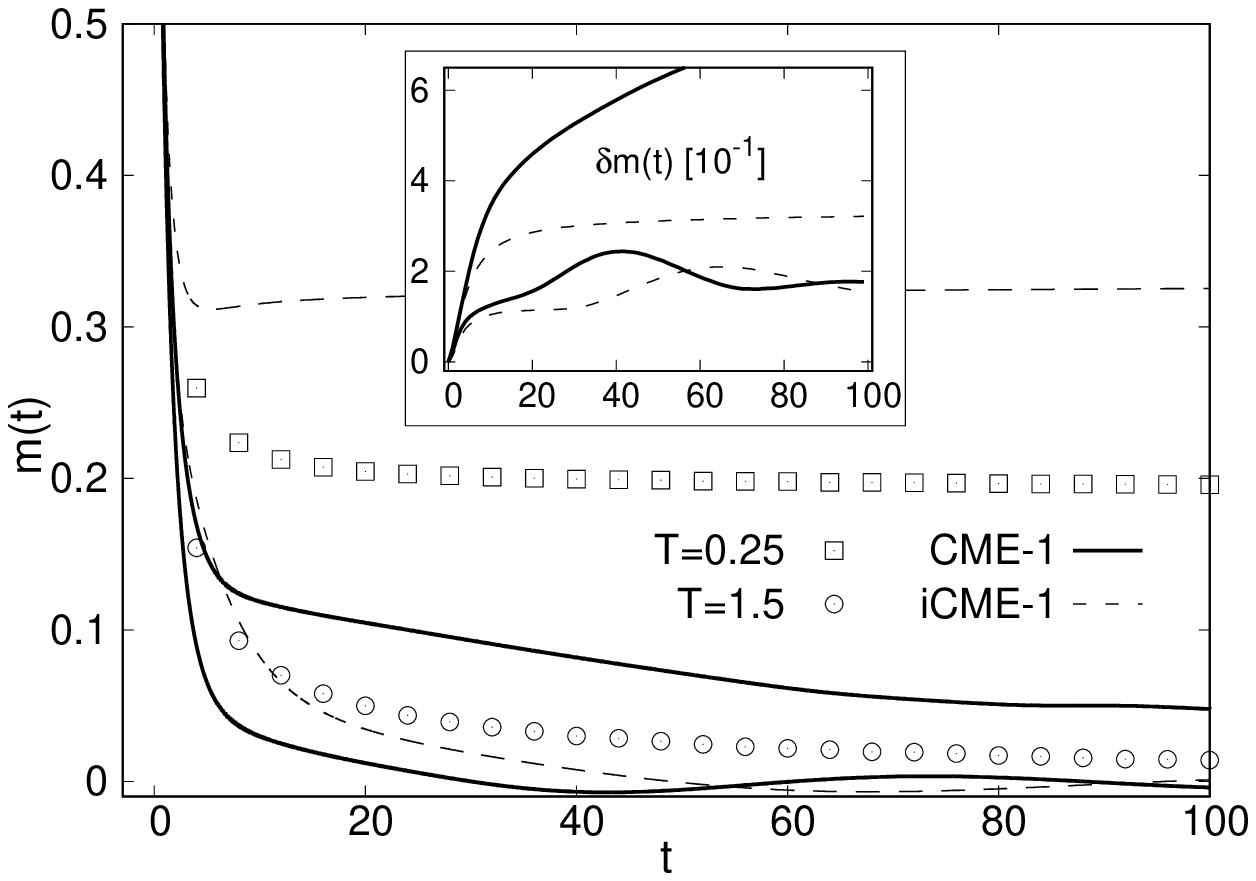}\label{fig:VB_mag}
}
\subfloat[]{
\includegraphics[width=0.35\textwidth]{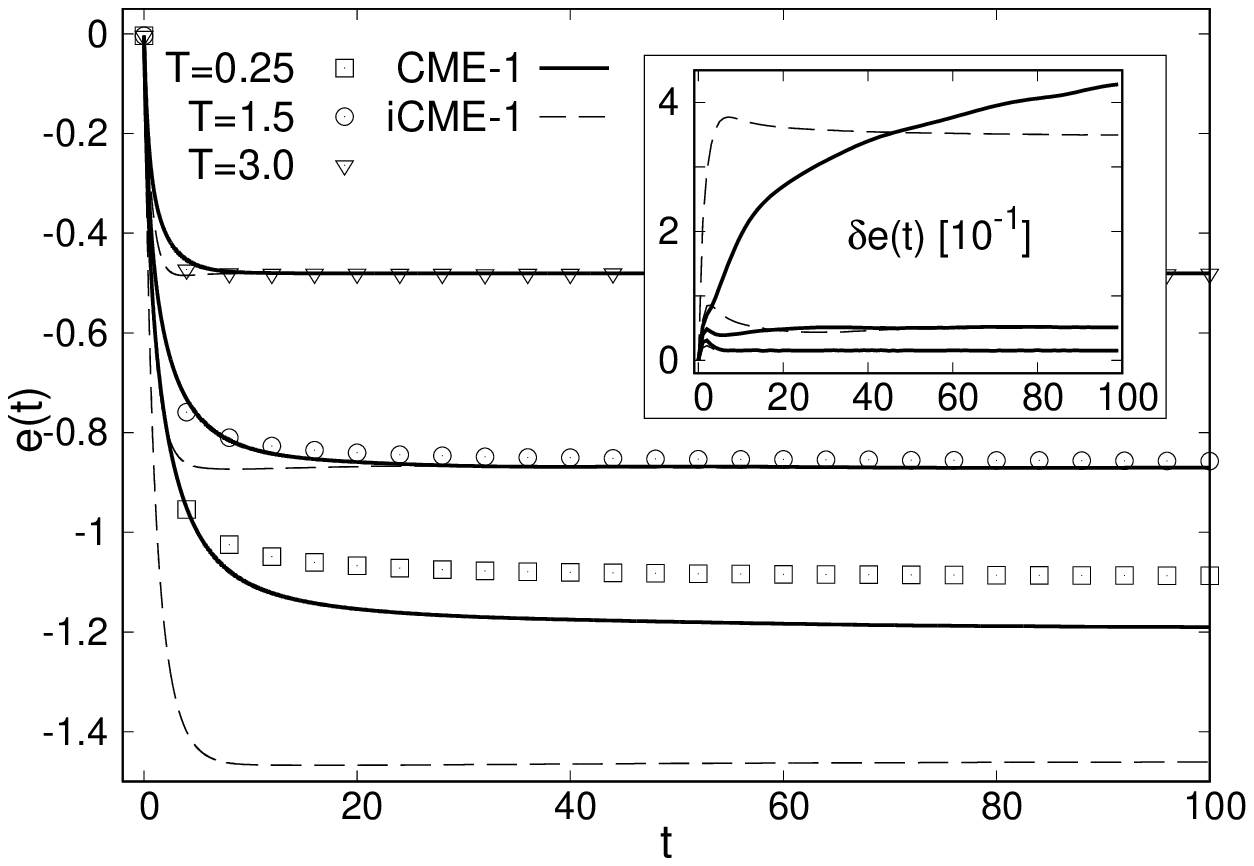}\label{fig:VB_ener}
}

\subfloat[]{
\includegraphics[width=0.35\textwidth]{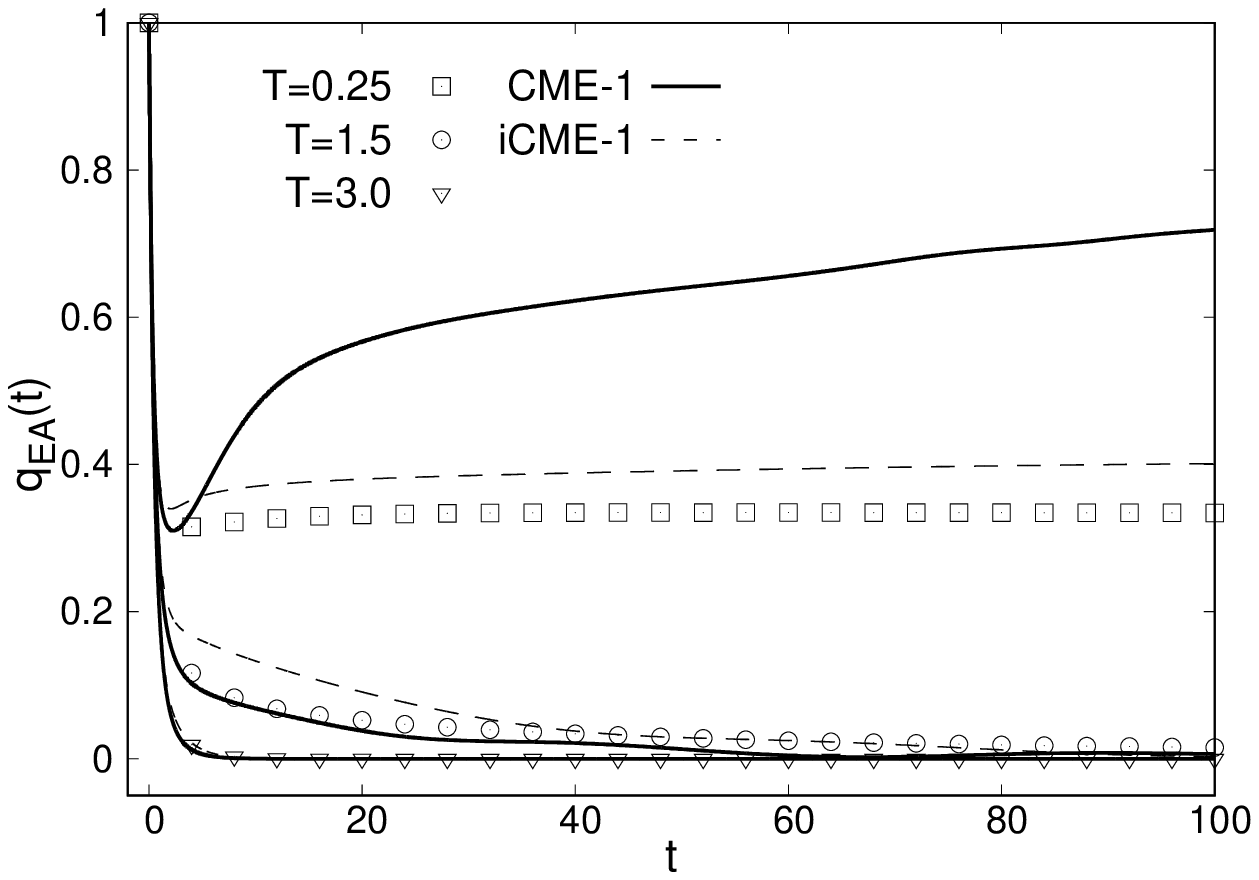}\label{fig:VB_qEA}
}
\caption{Continuous-time dynamics of Viana-Bray spin-glass model defined on a single instance of an Erdos-Renyi graph with $N=1000$ and average connectivity $c=3$. All panels compare the results of the earlier cavity closure \cite{CME-PRE} (thick continuous lines) and the new closure (dashed lines). As a reference, the results of Monte Carlo simulations of the dynamics are represented with points.
In all calculations, an initially fully magnetized system evolves in time in contact with a heat bath at a given temperature. Points are the average of $s=10,000$ Monte Carlo trajectories. \textbf{a)} Magnetization \textit{vs.} time. The inserted graphic shows the mean square error $\delta m(t) = (N^{-1}  \sum_{i=1}^{N} (m_i^{DCAV}(t) - m_i^{MC}(t))^{2})^{1/2}$ on scale of order $10^{-1}$. \textbf{b)} Energy density \textit{vs.} time. The inserted graphic shows the mean square error $\delta e(t) = (N^{-1} c^{-1}  \sum_{i\neq j} (e_{ij}^{DCAV}(t) - e_{ij}^{MC}(t))^{2})^{1/2}$ on scale of order $10^{-1}$.  
\textbf{c)} Edwards-Anderson parameter \textit{vs.} time.}
\label{fig:iCME_VB}
\end{figure}

So far, the integration of equations (\ref{eq:iCME-update}) and (\ref{eq:iCM-output}) outperforms the previous cavity theory, at least on these ferromagnetic systems. However, this is not necessarily true for all models. Let us take, for example, a case with a richer phenomenology, including an spin-glass phase for low temperatures. For this, we explored the Viana-Bray spin-glass model at low temperatures. As can be seen in Fig.~\ref{fig:iCME_VB} it is not so clear which cavity approach is better.

The main panel of Fig.~\ref{fig:VB_mag} shows the time evolution of the average magnetization. For a very low temperature ($T=0.25$ in the figure), all cavity theories are far from the Monte Carlo results, with no evident winner. The inserted graphic, on the other hand, gives smaller local errors $\delta m(t) = (N^{-1}  \sum_{i=1}^{N} (m_i^{DCAV}(t) - m_i^{MC}(t))^{2})^{1/2}$ for the new theory at the same low temperature. However, the time dependence of the average energy density is less trivial. Apparently, at low temperatures the new cavity method provides a much worse description than the equations derived in \cite{CME-PRE} (see Fig.~{\ref{fig:VB_ener}}). However, notice that, for the same temperatures the local error $\delta e(t) = (N^{-1} c^{-1}  \sum_{i\neq j} (e_{ij}^{DCAV}(t) - e_{ij}^{MC}(t))^{2})^{1/2}$ is smaller with the new approach. Finally, in Fig.~{\ref{fig:VB_qEA}} we show that the time evolution of the Edwards-Anderson parameter $q_{EA}$ is more precisely predicted by the new equations.

\section{Conclusions}
In summary, in this work we present a new scheme to close the Master Equation for a set of discrete random variables evolving in continuous time. The closure scheme is similar in spirit to the Dynamic Cavity method first introduced to study systems evolving in discrete time. We show that this closure scheme outperforms previous approximations describing the dynamics of KMC for the ferromagnetic Ising model. It is however, not good enough describing models with a glassy phase at low temperatures. In the direction to improve this and similar approaches we think that it is important to understand how to properly represent the history appearing in the equations of the Dynamical Cavity method. Moreover, it may turn fundamental to extend this schemes beyond the simple Replica Symmetry approximation. How to extend this theory to dynamics is not clear. On the technical level, iterations in 1-step Replica Symmetry Breaking (survey propagation) are weighted by a free energy shift. As non-equilibrium dynamics includes cyclic motion, in general it is not associated to a globally defined free energy function.

An alternative solution was proposed in \cite{Pelizzola,Gino2017} using a cluster variational method for a functional defined on the trajectories, one obtain equations that resemble (\ref{eq:cavity-update-X}), but where the graph of interaction takes into consideration the temporal correlation between the variables.

\section*{Acknowledgements}
We thank Profs Eduardo Dom\'{i}nguez and Federico Ricci-Tersenghi for numerous discussions. EA acknowledges support of the Swedish Research Council through grant 2020-04980. 

\appendix

\section{Derivation of Eq.~\protect\ref{eq:iCME-update}}
\label{app:derivation}
Following the definitions in \cite{Gino2015}, we will construct our cavity graph by taking a node $j$ and removing all its links but the one with its neighbor $i$. Thus, $j$ becomes an ``end node''. In that case one is able to give a closure for the master equation as we will show.
We denote the history of variable $j$ by $X_j$ and the history of variable $i$ by $X_i$.

Let us define the joint probability of $X_i$ and $X_j$ in the above defined cavity where $j$ is an end node:

\begin{equation}
 m_{i \to (ij)}^{t}(X_i, X_j) = \mu_{i \to (ij)}^{t} (X_i \: || \: X_j) \Phi^{t}(X_j \mid X_i) \label{eq:mess_pair}
\end{equation}
The notation $m_{i\to (ij)}$ indicates that this quantity
is a cavity probability ($m$), that it is a cavity probability
over histories $X_i$ and $X_j$ ($(ij)$), and that
it is so in the cavity of $j$ ($i\to $).

Now, let us find a differential equation for
the marginalization of $m_{i \to (ij)}^{t}(X_i, X_j)$
where only the dependence on the variable at the last
time instance is retained:

\begin{equation}
p_{i \to (ij)}^{t}(\sigma_i, \sigma_j) = \sum_{X_i \mid \sigma_i}^{t} \sum_{X_j \mid \sigma_j}^{t} m_{i \to (ij)}^{t}(X_i, X_j)
 \label{eq:pcav_pair}
\end{equation}

We need to expand the sum in the right hand side of (\ref{eq:pcav_pair}) to order $\Delta t$. More explicitly, we need to expand the sums:

\begin{eqnarray}
\sum_{X_i \mid \sigma_i}^{t+\Delta t} \sum_{X_j \mid \sigma_j}^{t+\Delta t} m_{i \to (ij)}^{t+\Delta t}(X_i, X_j) &=& \sum_{s_i} \int_{t_0}^{t+\Delta t} dt_1^{i} \ldots \int_{t_{s_i-1}^{i}}^{t+\Delta t} dt_{s_i}^{i} \sum_{s_j} \int_{t_0}^{t+\Delta t} dt_1^{j} \ldots \nonumber \\
& &  \ldots \int_{t_{s_j-1}^{j}}^{t+\Delta t} dt_{s_j}^{j} \: m_{i \to (ij)}^{t+\Delta t}(X_i, X_j) \label{eq:expl_sum}
\end{eqnarray}
where $s_i$ is the number of jumps in the history $X_i$, chosen such that the final state remains always $\sigma_i$, and $\{t_1^{i}, \ldots, t_{s_i}^{i}\}$ are the times at which these jumps occur. For the trajectory $X_j$, we analogously define the quantities $s_j$ and $\{t_1^{j}, \ldots, t_{s_j}^{j}\}$.

In the expansion, we need to keep only $O(\Delta t)$ terms. Thus, we can allow only two things:

\begin{itemize}
 \item[a)] All integrals are taken to time $t$, which means that no jumps occur between $t$ and $t+\Delta t$
 \item[b)] Only one integral is taken between $t$ and $t+\Delta t$, which means that only one jump occurs in that interval. The jump can correspond to $\sigma_i$ or $\sigma_j$
\end{itemize}

It is possible to parameterize $\mu_{i \to (ij)}^{t+\Delta t}(X_i \: || \: X_j)$ analogously to (\ref{eq:phi_param}), but using the cavity rates $\lambda^{t}_{i \to (ij)}(X_i, X_j)$. This $\lambda$ represents the probability per time unit of having a jump in the trajectory $X_i$ at time $t$, with the information that $\sigma_j$ has followed some trajectory $X_j$.

When no jumps occur we have:

\begin{eqnarray}
 \mu_{i \to (ij)}^{t+\Delta t}(X_i \: || \: X_j) &=& \big[1 - \lambda^{t}(X_i, X_j) \, \Delta t \big] \; \mu_{i \to (ij)}^{t}(X_i \: || \: X_j) + o(\Delta t) \label{eq:exp_mu_no_jumps} \\
 \Phi^{t+\Delta t}(X_j \mid X_i) &=& \big[1 - r(\sigma_j, \sigma_i)  \, \Delta t \big] \; \Phi^{t}(X_j \mid X_i) + o(\Delta t) \label{eq:exp_phi_no_jumps}
\end{eqnarray}
where we used the shorthand $\lambda^{t}(X_i, X_j)$ for the cavity rates.

Substituting (\ref{eq:exp_mu_no_jumps}) and (\ref{eq:exp_phi_no_jumps}) into (\ref{eq:expl_sum}) we get the first contribution:

\begin{eqnarray}
 I_0 &=& p_{i \to (ij)}^{t}(\sigma_i, \sigma_j) - \Delta t \: r(\sigma_j, \sigma_i)  p_{i \to (ij)}^{t}(\sigma_i, \sigma_j) - \nonumber \\
 & & \:\:\: - \Delta t \sum_{X_i \mid \sigma_i}^{t} \sum_{X_j \mid \sigma_j}^{t} \lambda^{t}(X_i, X_j) \, m_{i \to (ij)}^{t}(X_i, X_j) \label{eq:I0}
\end{eqnarray}

On the other hand, what happens when the last jumps occurs between $t$ and $t+\Delta t$ is a little different. Let us
consider first $\Phi^{t+\Delta t}(X_j \mid X_i)$, which can be parameterized as in equation (\ref{eq:phi_param}).
When the last jump in $X_j$ occurs at $t_{s_j} \in (t, t+\Delta t)$, and taking into account that $\sigma_i$ will not jump in that interval, we can write:

\begin{eqnarray}
\Phi^{t+\Delta t}(X_j \mid X_i) = \Phi^{t}(X_j^{-} \mid X_i) \: r_j(-\sigma_j, \sigma_i) \: e^{- (t_{s_j} - t) \, r(-\sigma_j, \sigma_i)} \times \nonumber \\  \:\:\:\:\:\:\:\:\:\:\:\:\:\:\:\:\:\:\:\:\:\:\:\:\:\:\:\:\:\:\:\:\:\:\:\:\:\:\:\:\:\:\:\:\:\:\:\:\:\:\:\:\:\:\:\:\:\:\:\:\:\:\:\:\:\:\:\:\:\:\:\:\:\:\:\:\:\:\:\: \times \:\: e^{-(t+\Delta t - t_{s_j} ) \, r(\sigma_j, \sigma_i)}
 \label{eq:phi_param_last_jump}
\end{eqnarray}
where $X_j^{-}$ is a trajectory that ends up with the value $X_j^{-}(t)=-\sigma_j$.

As the expression (\ref{eq:phi_param_last_jump}) will be inside an integral that is already of order $\Delta t$, we can keep only the order zero terms:

\begin{eqnarray}
\Phi^{t+\Delta t}(X_j \mid X_i) = \Phi^{t}(X_j^{-} \mid X_i) \: r_j(-\sigma_j, \sigma_i) + O(\Delta t)
 \label{eq:phi_param_last_jump_or_zero}
\end{eqnarray}

The second contribution is, thus:

\begin{eqnarray}
 I_1 &=& \Delta t \: r(-\sigma_j, \sigma_i)  p_{i \to (ij)}^{t}(\sigma_i, -\sigma_j)\label{eq:I1}
\end{eqnarray}

Analogously, in the case where only $\sigma_i$ jumps in the interval $(t, t + \Delta t)$, we have:

\begin{eqnarray}
 I_2 &=& \Delta t \sum_{X_i^{-} \mid -\sigma_i}^{t} \sum_{X_j \mid \sigma_j}^{t} \lambda^{t}(X_i^{-}, X_j) \, m_{i \to (ij)}^{t}(X_i^{-}, X_j)\label{eq:I2}
 \end{eqnarray}

As can be seen from (\ref{eq:I0}) and (\ref{eq:I2}), we still need to eliminate the cavity rates $\lambda^{t}$ from our equations, because we do not know its exact form. Nevertheless, in analogy to the derivation provided in \cite{CME-PRE}, we can use the message-passing equation (\ref{eq:cavity-update-X}) to write:

\begin{eqnarray}
 \! \! \lambda^{t}(X_i, X_j) \, m_{i \to (ij)}(X_i ,\!  X_j)  = \!\!  \sum_{\sigma_{\partial i \setminus j}} r_{i}(\sigma_i, \sigma_{\partial i \setminus j}, \sigma_j) \, p_{i \to (ij)}^{t}(\sigma_{\partial i \setminus j}, X_i, X_j) \label{eq:lambda_m}
\end{eqnarray}

With this, it is easy to rewrite the contributions (\ref{eq:I0}) and (\ref{eq:I2}). Putting the results together with (\ref{eq:I1}), we obtain a new tree-exact CME for the pair probability densities in equation (\ref{eq:pcav_pair}):

\begin{eqnarray}
\frac{d p_{i \to (ij)}(\sigma_i, \sigma_j)}{dt}  &=&  -\sum_{\sigma_{\partial i \setminus j}} r_i(\sigma_i, \sigma_{\partial i}) \, p_{i \to (ij)}(\sigma_{\partial i \setminus j}, \sigma_i, \sigma_j) + \nonumber \\
& & + \sum_{\sigma_{\partial i \setminus j}} r_i(-\sigma_i, \sigma_{\partial i}) \, p_{i \to (ij)}(\sigma_{\partial i \setminus j}, -\sigma_i, \sigma_j) - \nonumber \\
& &  - r_j(\sigma_j, \sigma_i) \, p_{i \to (ij)}^{t}(\sigma_i, \sigma_j) + \nonumber \\
& & +  r_j(-\sigma_j, \sigma_i) \, p_{i \to (ij)}^{t}(\sigma_i, -\sigma_j)
 \label{eq:iCME_tree_exact}
\end{eqnarray}

\section*{Bibliography}
\bibliographystyle{unsrt}
\bibliography{references,references2}

\end{document}